\documentstyle[aps,pra]{revtex}
\begin{document}
\draft

\newcommand{\uu}[1]{\underline{#1}}
\newcommand{\pp}[1]{\phantom{#1}}
\newcommand{\be}{\begin{eqnarray}}
\newcommand{\ee}{\end{eqnarray}}
\newcommand{\ve}{\varepsilon}
\newcommand{\vs}{\varsigma}
\newcommand{\Tr}{{\,\rm Tr\,}}
\newcommand{\pol}{\frac{1}{2}}

\title{
Non-canonical quantization of electromagnetic fields 
and the meaning of $Z_3$
}
\author{Marek~Czachor}
\address{
Katedra Fizyki Teoretycznej i Metod Matematycznych\\
Politechnika Gda\'{n}ska,
ul. Narutowicza 11/12, 80-952 Gda\'{n}sk, Poland}
\maketitle

\begin{abstract}
Non-canonical quantization is based on certain reducible
representations of canonical commutation relations. 
Relativistic formalism for electromagnetic 
non-canonical quantum fields is introduced. Unitary
representations of the Poincar\'e group at the level of fields and
states are explicitly given. Multi-photon and coherent states are introduced.
Statistics of photons in a coherent state is Poissonian if an
appropriately defined thermodynamic limit is performed. Radiation
fields having a correct $S$ matrix are constructed. The $S$ matrix is
given by a non-canonical coherent-state displacement operator, a fact
automatically eliminating the infrared catastrophe.  
This, together with earlier results on elimination of vacuum and
ultraviolet infinities, suggests that non-canonical quantization
leads to finite field theories. Renormalization constant $Z_3$ is
identified with a parameter related to wave functions of non-canonical
vacua. 
\end{abstract}
%\narrowtext

\section{Introduction}

The idea
of non-canonical quantization of electromagnetic fields was introduced in
\cite{I}. A starting point is an observation that even in
nonrelativistic quantum mechanics it is natural to treat the
frequency $\omega$ characterizing a harmonic oscillator as an
eigenvalue and not a parameter. A detailed analysis of a
nonrelativistic indefinite-frequency harmonic oscillator given in
\cite{II} allowed to 
better understand physical structures associated with this
modification. The main
results obtained so far were

(a) Association of an entire spectrum of frequencies with a single
indefinite-frequency oscillator.

(b)
Non-canonical algebra of creation and annihilation operators. 

(c)
An ensemble of $N$ indefinite-frequency oscillators having properties of
a non-canonical quantum field. 

(d)
Finite vacuum energy and finite vacuum fluctuations.

(e)
Structures of the canonical quantum field theory obtained 
in the thermodynamic limit $N\to \infty$, but with form-factors
regularizing ultraviolet infinities.

(f)
The form-factors appearing automatically as a consequence of
noncanonical commutation relations and the structure of vacuum, and
not introduced in an ad hoc manner.  

(g)
Space of states in a form of a vector bundle with the set of vacua in
a role of a base space and Fock-type fibers. 
Probabilities have to be associated with a single fiber. In particular, the vacuum is in a given fiber the unique vector obtained by projecting 
the fiber on the base space. 

In \cite{I} it was also concluded that non-canonical quantization may
lead to deviations from the Planck black-body law. However, a
consistent interpretation of the thermodynamic limit shows that no
deviations should be expected (see below). 

In the present paper we extend the formalism to the relativistic
domain. We begin with quantization in free space. Representation of
non-canonical commutation relations and the Poincar\'e group are
explicitly constructed.  
Multi-photon and coherent states are defined. Their properties make
them similar to those from the canonical formalism if one performs a
thermodynamic limit. Vacuum states have properties of
translation-invariant scalar fields and are Poincar\'e covariant. 
Commutation relations for vector potentials are
found showing certain deviations from locality due to nontrivial
structure of vacua.
Radiation fields leading to the correct form of
the $S$ matrix are found, the $S$ matrix being proportional to a
non-canonical coherent-state displacement operator. The same
mechanism that regularized vacuum and ultraviolet infinities is shown
to regularize the infrared divergence. Finally, we identify the
renormalization constant $Z_3$ with an amplitude related to the
electromagnetic vacuum wave function. The latter result is fully consistent 
with the analysis of perturbation theory given in \cite{I} and
\cite{II}. 

All the results obtained so far lead to the same conclusion: The main
difficulty of contemporary quantum field theories is rooted in the
fact that they are {\it too classical\/}. The reinterpretation of
some parameters in terms of eigenvalues simultaneously removes 
the ultraviolet, infrared, and vacuum infinities, and explains the
origin of renormalization constants. The above conclusions are valid,
strictly speaking, for bosons. A work on fermions is in progress.

\section{Notation}

In order to control covariance properties of fields in generalized
frameworks it is best to work in a manifestly covariant formalism.
The most convenient is the one based on spinors and passive unitary
transformations. 

\subsection{Spinor convention and fields}

We take $c=1$ and $\hbar=1$.
The index notation we use in the paper is consistent with the
Penrose-Rindler spinor and world-tensor convention \cite{PR}. The
electromagnetic field-tensor and its dual are 
\be
F_{ab}
&=&
\left(
\begin{array}{cccc}
0    & E_1 & E_2  & E_3\\
-E_1 & 0   & -B_3 & B_2 \\
-E_2 & B_3 & 0 & -B_1\\
-E_3 & -B_2 & B_1 & 0 
\end{array}
\right)\\
^*F_{ab}
&=&
\left(
\begin{array}{cccc}
0    & -B_1 & -B_2  & -B_3\\
B_1 & 0   & -E_3 & E_2 \\
B_2 & E_3 & 0 & -E_1\\
B_3 & -E_2 & E_1 & 0 
\end{array}
\right)
\ee
Self-dual and anti-self-dual parts of $F_{ab}$ are related to the
electromagnetic spinor by 
\be
^+ F_{ab}
&=&
\frac{1}{2}
\Big(F_{ab}-i ^* F_{ab}\Big)=\varepsilon_{AB}\bar \varphi_{A'B'}
=
\left(
\begin{array}{cccc}
0    & F_1 & F_2  & F_3\\
-F_1 & 0   & iF_3 & -iF_2 \\
-F_2 & -iF_3 & 0 & iF_1\\
-F_3 & iF_2 & -iF_1 & 0 
\end{array}
\right)\\
^- F_{ab}
&=&
\frac{1}{2}
\Big(F_{ab}+i ^* F_{ab}\Big)=\varepsilon_{A'B'}\varphi_{AB}
\ee
where $\bbox F=(\bbox E+i\bbox B)/2$ is the Riemann-Silberstein
vector \cite{IBB-pwf}. Denote $k\cdot x=k_ax^a$.  
The electromagnetic spinor has the following Fourier representation
\cite{Woodhouse,IZBB}
\be
\varphi_{AB}(x)
&=&
\int d\Gamma(\bbox k)\pi_{A}(\bbox k)\pi_{B}(\bbox k)
\Big(f(\bbox k,-)e^{-ik\cdot x}+\overline{f(\bbox k,+)}e^{ik\cdot
x}\Big)\label{varphi}\\ 
\bar\varphi_{A'B'}(x)
&=&
\int d\Gamma(\bbox k)\bar\pi_{A'}(\bbox k)\bar\pi_{B'}(\bbox k)
\Big(f(\bbox k,+)e^{-ik\cdot x}
+\overline{f(\bbox k,-)}e^{ik\cdot x}\Big)\label{varphi'}
\ee
where the spinor field $\pi_{A}(\bbox k)$ is related to
the future-pointing 4-momentum by 
\be
k^a&=& \pi^{A}(\bbox k)\bar \pi^{A'}(\bbox k)=(k_0,\bbox k)
=(|\bbox k|,\bbox k)
\ee
and the invariant measure on the light-cone is $d\Gamma(\bbox k)=
\big[(2\pi)^3 2k_0\big]^{-1}d^3k$.
Anti-self-dual and self-dual parts of the field tensor are 
\be
^- F_{ab}(x)
&=&
\int d\Gamma(\bbox k)\varepsilon_{A'B'}\pi_{A}(\bbox k)\pi_{B}(\bbox k)
\Big(f(\bbox k,-)e^{-ik\cdot x}
+\overline{f(\bbox k,+)}e^{ik\cdot x}\Big)\\
&=&
\int d\Gamma(\bbox k)e_{ab}(\bbox k)
\Big(f(\bbox k,-)e^{-ik\cdot x}
+\overline{f(\bbox k,+)}e^{ik\cdot x}\Big)\\
^+ F_{ab}(x)
&=&
\int d\Gamma(\bbox k)\varepsilon_{AB}\bar\pi_{A'}(\bbox
k)\bar\pi_{B'}(\bbox k) 
\Big(\overline{f(\bbox k,-)}e^{ik\cdot x}
+f(\bbox k,+)e^{-ik\cdot x}\Big)\\
&=&
\int d\Gamma(\bbox k)\bar e_{ab}(\bbox k)
\Big(\overline{f(\bbox k,-)}e^{ik\cdot x}
+f(\bbox k,+)e^{-ik\cdot x}\Big)
\ee
The latter form is used by the Bia{\l}ynicki-Birulas in \cite{IZBB}.

The sign in the amplitude $f(\bbox k,\pm)$ corresponds to the value
of helicity of positive-frequency fields. 

The four-vector potential $A_a(x)$ is related to the electromagnetic spinor by
\be
\varphi_{XY}(x)
&=&
\nabla{_{(X}}{^{Y'}}A{_{Y)Y'}}(x)\label{A}
\ee
In the Lorenz gauge $\nabla^aA_a=0$ we do not have to symmetrize the unprimed
indices and
\be
\varphi_{XY}(x)
&=&
\nabla{_{X}}{^{Y'}}A{_{YY'}}(x).
\ee
One of the possible Lorenz gauges is 
\be
A_a(x)
&=&
i\int d\Gamma(\bbox k)\Bigg(
m_{a}(\bbox k)
\Big(f(\bbox k,+)e^{-ik\cdot x}-\overline{f(\bbox k,-)}e^{ik\cdot x}\Big)
-
\bar m_{a}(\bbox k)
\Big(\overline{f(\bbox k,+)}e^{ik\cdot x}-f(\bbox k,-)e^{-ik\cdot x}\Big)
\Bigg)
\label{gauge}
\ee
where $\omega_A\pi^A=1$, i.e $\omega_A=\omega_A(\bbox k)$ is a
spin-frame partner of $\pi^A(\bbox k)$.  
In (\ref{gauge}) we have introduced the null vectors 
\be
m_{a}(\bbox k)
&=&
\omega_A(\bbox k)\bar\pi_{A'}(\bbox k)\\
\bar m_{a}(\bbox k)
&=&
\pi_A(\bbox k)\bar\omega_{A'}(\bbox k)
\ee
which, together with 
\be
k_a &=&
\pi_A(\bbox k)\bar\pi_{A'}(\bbox k)\\
\omega_a(\bbox k)
&=&
\omega_A(\bbox k)\bar\omega_{A'}(\bbox k)
\ee
form a null tetrad \cite{PR}. 

A change of gauge is in the
Fourier domain represented by a shift by 
a multiple of $k^a$. The form (\ref{gauge}) shows that gauge freedom
is related to the nonuniqueness of $\omega_A(\bbox k)$ which can be
shited by a multiple of $\pi_A(\bbox k)$.

\subsection{Momentum representation}

Consider the momentum-space basis normalized by
\be
\langle \bbox p|\bbox p'\rangle=(2\pi)^3 2p_0\delta^{(3)}(\bbox
p-\bbox p') =\delta_\Gamma(\bbox p,\bbox p')
\ee
The identity operator in momentum space is
$\int d\Gamma(\bbox p)|\bbox p\rangle\langle \bbox p|$. 
We can use the following explicit realization of 
$|\bbox p\rangle=|f_p\rangle$ in terms of distributions
\be
f_p(\bbox k)
&=&
(2\pi)^32 p_0 \delta^{(3)}(\bbox p-\bbox k)
=
\delta_\Gamma(\bbox p,\bbox k)
\ee
Since 
\be
\int d\Gamma(\bbox k) F(\bbox k)\delta_\Gamma(\bbox p,\bbox k)
&=&
\int d^3k \delta^{(3)}(\bbox p-\bbox k)F(\bbox k)=F(\bbox p)
\ee
the Fourier transform of $f_p(\bbox k)$ is 
\be
\check f_p(x)
&=&
\int d\Gamma(\bbox k)f_p(\bbox k) e^{-i k\cdot x}=
e^{-i p\cdot x}
\ee
If $1$ is the identity operator occuring at the right-hand-side of CCR 
$[a_s,a^{\dag}_{s'}]=\delta_{ss'}1$, we denote 
\be
I_{\bbox k}
&=&
|\bbox k\rangle\langle \bbox k|\otimes 1,\\
I
&=&
\int d\Gamma(\bbox k)|\bbox k\rangle\langle \bbox k|\otimes 1.
\ee

\subsection{Multi-particle conventions}

Let $A$ be an operator $A: {\cal H}\to {\cal H}$ where $\cal H$ is a
one-particle Hilbert space. The multi-particle Hilbert space 
\be
\uu {\cal H}=\oplus_{n=1}^\infty \otimes_s^n {\cal H}
\ee
is the Hilbert space of states corresponding to an indefinite number
of bosonic particles; $\otimes_s^n {\cal H}$ stands for a space of
symmetric states in $\underbrace{{\cal H}\otimes\dots \otimes {\cal H}}_n$.
We introduce the following notation for operators defined at the
multi-particle level:
\be
\oplus_{\alpha_n} A
&=&
\alpha_1 A
\oplus 
\alpha_2\big(A\otimes I+I\otimes A\big)\oplus
\alpha_3\big(A\otimes I\otimes I+
I\otimes A \otimes I+
I\otimes I\otimes A\big)\oplus
\dots
\ee
Here $\oplus_{\alpha_n} A: \uu {\cal H}\to \uu {\cal H}$,
$\alpha_n$ are real or complex parameters, and $I$ is the identity
operator in $\cal H$.  

The following properties follow directly from the definition 
\be
[\oplus_{\alpha_n} A,\oplus_{\beta_n} B]&=&\oplus_{\alpha_n\beta_n}
[A,B]\\
e^{\oplus_{\alpha_n} A}
&=&
\oplus_{n=1}^\infty
\underbrace{e^{\alpha_n A}\otimes\dots \otimes e^{\alpha_n A}}_n\\
e^{\oplus_1 A}\oplus_{\beta_n} Be^{-\oplus_1 A}
&=&
\oplus_{\beta_n}e^{A}Be^{-A}
\ee
Identity operators  in $\uu{\cal H}$ and $\cal H$ are related by 
\be
\uu I &=& \oplus_{\frac{1}{n}} I.
\ee 
We will often use the operator
\be
\uu I_{\bbox k}
&=&
\oplus_{\frac{1}{n}} I_{\bbox k}.
\ee

\section{Poincar\'e transformations of classical electromagnetic fields}

Denote, respectively, by $\Lambda$ and $y$ the $SL(2,C)$ and
4-translation parts of a Poincar\'e transformation\cite{PT}
$(\Lambda,y)$.  
The spinor representation of the Poincar\'e group acts in the space
of anti-self-dual electromagnetic fields in 4-position representation
as follows: 
\be
^-\hat F_{ab}(x)
&\mapsto&
\big(T_{\Lambda,y}{^-}\hat F\big)_{ab}(x)\\
&=&
\Lambda{_a}{^c}\Lambda{_b}{^d}{^-}\hat F_{cd}\big(\Lambda^{-1}(x-y)\big)\\
&=&
\int d\Gamma(\bbox k)\varepsilon_{A'B'}\Lambda{_A}{^C}\pi_{C}(\bbox k)
\Lambda{_B}{^D}\pi_{D}(\bbox k)
\Big(f(\bbox k,-)e^{-ik\cdot \Lambda^{-1}(x-y)}
+\overline{f(\bbox k,+)}e^{ik\cdot \Lambda^{-1}(x-y)}\Big)\\
&=&
\int d\Gamma(\bbox k)\varepsilon_{A'B'}\Lambda{_A}{^C}\pi_{C}(\bbox
{\Lambda^{-1}k}) 
\Lambda{_B}{^D}\pi_{D}(\bbox {\Lambda^{-1}k})
\Big(f(\bbox {\Lambda^{-1}k},-)e^{-ik\cdot (x-y)}
+\overline{f(\bbox {\Lambda^{-1}k},+)}e^{ik\cdot (x-y)}\Big)
\ee
where $\bbox{\Lambda^{-1}k}$ is the spacelike part of
$\Lambda^{-1}{_a}{^b}k{_b}$.  
The transformed field 
\be
(\Lambda\pi){_{A}}(\bbox k)
=
\Lambda{_A}{^C}\pi_{C}(\bbox {\Lambda^{-1}k})
\ee
satisfies 
\be
k^a&=& \pi^{A}(\bbox k)\bar \pi^{A'}(\bbox k)
=
(\Lambda\pi){^{A}}(\bbox k)\overline{\Lambda\pi}{^{A'}}(\bbox k)
\ee
Now, if $\omega_{A}(\bbox k)$ is a spin-frame partner of 
$\pi_{A}(\bbox k)$, i.e. $\omega_{A}(\bbox k)\pi^{A}(\bbox k)=1$, 
one can write 
\be
\pi^{A}(\bbox k)
&=&
(\Lambda\pi){^{A}}(\bbox k)\bar \omega_{A'}(\bbox
k)\overline{\Lambda\pi}{^{A'}}(\bbox k) 
\ee
which shows that 
$\pi^{A}(\bbox k)$ and $(\Lambda\pi){_{A}}(\bbox k)
=
\Lambda{_A}{^C}\pi_{C}(\bbox {\Lambda^{-1}k})$
are proportional to each other, the proportionality factor being 
\be
\lambda(\Lambda,\bbox k)
=\bar \omega_{A'}(\bbox
k)\overline{\Lambda\pi}{^{A'}}(\bbox k).
\ee
The form (\ref{gauge}) showed that the gauge freedom is related to
shifts 
\be
\omega_A(\bbox k)\mapsto \omega_A(\bbox k)+{\rm scalar}\times 
\pi_A(\bbox k)
\ee
which do not affect $\lambda(\Lambda,\bbox k)$ making it independent
of gauge. 
Using again 
\be
k^a&=&|\lambda(\Lambda,\bbox k)|^2 \pi^{A}(\bbox k)\bar \pi^{A'}(\bbox k)
=|\lambda(\Lambda,\bbox k)|^2k^a
\ee
one concludes that $\lambda(\Lambda,\bbox k)$ is a phase factor
\be
\lambda(\Lambda,\bbox k)=e^{i\Theta(\Lambda,\bbox k)}
\ee
and we find
\be
\big(T_{\Lambda,y}{^-}\hat F\big)_{ab}(x)
&=&
\int d\Gamma(\bbox k)\varepsilon_{A'B'}\pi_{A}(\bbox {k})
\pi_{B}(\bbox {k})e^{-2i\Theta(\Lambda,\bbox k)}
\Big(f(\bbox {\Lambda^{-1}k},-)e^{-ik\cdot (x-y)}
+\overline{f(\bbox {\Lambda^{-1}k},+)}e^{ik\cdot (x-y)}\Big)\\
&=&
\int d\Gamma(\bbox k)
e_{ab}(\bbox {k})
\Big(e^{-2i\Theta(\Lambda,\bbox k)}e^{ik\cdot y}
f(\bbox {\Lambda^{-1}k},-)e^{-ik\cdot x}
+
e^{-2i\Theta(\Lambda,\bbox k)}e^{-ik\cdot y}
\overline{f(\bbox {\Lambda^{-1}k},+)}e^{ik\cdot x}\Big)\\
&=&
\int d\Gamma(\bbox k)
e_{ab}(\bbox {k})
\Big(
(T_{\Lambda,y}f)(\bbox {k},-)e^{-ik\cdot x}
+
\overline{(T_{\Lambda,y}f)(\bbox {k},+)}e^{ik\cdot x}\Big)
\ee
We have obtained therefore the {\it passive\/} transformation of the
classical wave function 
\be
f(\bbox {k},\pm)
\mapsto
(T_{\Lambda,y}f)(\bbox {k},\pm)
=
e^{\pm 2i\Theta(\Lambda,\bbox k)}e^{ik\cdot y}
f(\bbox {\Lambda^{-1}k},\pm).\label{unit}
\ee
which is simply the unitary zero-mass spin-1 representation 
the Poincar\'e group. The above derivation clearly shows that the
rule (\ref{unit}) is obtained without any particular assumption about
the choice of $f(\bbox {k},\pm)$. In particular, the derivation
remains valid even if one replaces functions $f$ by operators,
independently of their algebraic properties.  
The above `passive' viewpoint on the structure of unitary
representations is particularly useful if one aims at generalizations
of CCR. The passive derivation of all the non-tachyonic unitary
representations of the Poincar\'e group can be found in \cite{MC-BW}.

\section{Non-canonical quantization}

We follow the strategy described in \cite{I} and \cite{II}. Let $a_s$
be {\it canonical\/} annihilation operators satisfying CCR
$[a_s,a^{\dag}_{s'}]=\delta_{ss'}1$.  
Define the 1-oscillator {\it non-canonical\/} creation and
annihilation operators \cite{uwaga} 
\be
a(f)^{\dag}
&=&
\sum_s\int d\Gamma(\bbox k)f(\bbox k,s)|\bbox k\rangle\langle \bbox
k| \otimes a^{\dag}_s\\
&=&
\sum_s\int d\Gamma(\bbox k)f(\bbox k,s)a(\bbox k,s)^{\dag}\label{a(f)}\\
a(f)
&=&
\sum_s\int d\Gamma(\bbox k)\overline{f(\bbox k,s)}
|\bbox k\rangle\langle \bbox k| \otimes a_s\\
&=&
\sum_s\int d\Gamma(\bbox k)\overline{f(\bbox k,s)}
a(\bbox k,s)\label{a(f)^+}
\ee
satisfying the non-CCR algebra
\be
[a(\bbox k,s),a(\bbox k',s')^{\dag}]
&=&
\delta_{ss'}\delta_\Gamma(\bbox k, \bbox k')
|\bbox k\rangle\langle \bbox k| \otimes 1\\
&=&
\delta_{ss'}\delta_\Gamma(\bbox k, \bbox k')
I_{\bbox k}.
\ee 
Taking, in particular, $f_{p,r}(\bbox k,s)=\delta_{rs}
\delta_\Gamma(\bbox p,\bbox k)$ one finds 
\be
a(f_{p,r})
&=&
|\bbox p\rangle\langle \bbox p| \otimes a_r
=
a(\bbox p,r).\label{53}
\ee
The one-oscillator quantization is 
\be
^-\hat F_{ab}(x)
&=&
\varepsilon_{A'B'}\hat \varphi_{AB}(x)\\
&=&
\int d\Gamma(\bbox k)\varepsilon_{A'B'}\pi_{A}(\bbox k)\pi_{B}(\bbox k)
\Big(a(\bbox k,-)e^{-ik\cdot x}
+a(\bbox k,+)^{\dag}e^{ik\cdot x}\Big)\\
&=&
\int d\Gamma(\bbox k)e_{ab}(\bbox k)
\Big(a(\bbox k,-)e^{-ik\cdot x}
+a(\bbox k,+)^{\dag}e^{ik\cdot x}\Big)
\ee
Spinor transformations of $^-\hat F_{ab}(x)$ lead to the passive
transformation 
\be
a(\bbox {k},\pm)
\mapsto
(T_{\Lambda,y}a)(\bbox {k},\pm)
=
e^{\pm 2i\Theta(\Lambda,\bbox k)}e^{ik\cdot y}
a(\bbox {\Lambda^{-1}k},\pm).
\ee
The quantization procedure is gauge independent since we work at the
gauge-independent level of $^-\hat F_{ab}(x)$.  

Multi-oscillator fields are defined in terms of 
\be
\uu a(\bbox p,s)
&=&
\oplus_\frac{1}{\sqrt{n}}
a(\bbox p,s)\label{58}
\ee
and 
\be
\uu a(f)^{\dag}
&=&
\sum_s\int d\Gamma(\bbox k)f(\bbox k,s)\uu a(\bbox k,s)^{\dag}\\
\uu a(f)
&=&
\sum_s\int d\Gamma(\bbox k)\overline{f(\bbox k,s)}
\uu a(\bbox k,s).
\ee
The non-CCR algebra is
\be
[\uu a(f),\uu a(g)^{\dag}]
&=&
\sum_{s}\int d\Gamma(\bbox k)
\overline{f(\bbox k,s)}g(\bbox k,s)
\uu I_{\bbox k}
\ee
The right-hand-side of the above formula is in the center of the
non-CCR algebra, i.e. 
\be
\big[[\uu a(f),\uu a(g)^{\dag}],\uu a(h)\big]
&=&
0\\
\big[[\uu a(f),\uu a(g)^{\dag}],\uu a(h)^{\dag}\big]
&=&
0
\ee
Useful is also the formula 
\be
[\uu a(f_{p,r}),\uu a(f_{p',r'})^{\dag}]
&=&
[\uu a(\bbox p,r),\uu a(\bbox p',r')^{\dag}]\\
&=&
\delta_{rr'}
\delta_\Gamma(\bbox p,\bbox p')\uu I_{\bbox p}.\label{65}
\ee
The presence of $\uu I_{\bbox p}$ at the right-hand-sides of
non-CCR will influence orthogonality properties of multi-photon
states, as we shall see later.  

At the multi-oscillator level the electromagnetic field tensor operator is 
\be
{\uu F}_{ab}(x)
&=&
\int d\Gamma(\bbox k)e_{ab}(\bbox k)
\Big(\uu a(\bbox k,-)e^{-ik\cdot x}
+\uu a(\bbox k,+)^{\dag}e^{ik\cdot x}\Big)\\
&+&
\int d\Gamma(\bbox k)\bar e_{ab}(\bbox k)
\Big(\uu a(\bbox k,-)^{\dag}e^{ik\cdot x}
+\uu a(\bbox k,+)e^{-ik\cdot x}\Big).
\ee
The four-potential operator is (in our gauge)
\be
{\uu A}_a(x)
&=&
i\int d\Gamma(\bbox k)\Bigg(
m_{a}(\bbox k)
\Big(\uu a(\bbox k,+)e^{-ik\cdot x}
-\uu a(\bbox k,-)^{\dag}e^{ik\cdot x}\Big)
+
\bar m_{a}(\bbox k)
\Big(\uu a(\bbox k,-)e^{-ik\cdot x}
-\uu a(\bbox k,+)^{\dag}e^{ik\cdot x}\Big)
\Bigg).
\ee

\section{Action of the Poincar\'e group on field operators}

We are interested in finding the representation of the group in terms
of unitary 
similarity transformations, i.e.
\be
\uu a(\bbox {k},\pm)
\mapsto
e^{\pm 2i\Theta(\Lambda,\bbox k)}e^{ik\cdot y}
\uu a(\bbox {\Lambda^{-1}k},\pm)
=
{\uu U}_{\Lambda,y}^{\dag} \uu a(\bbox {k},\pm){\uu U}_{\Lambda,y}
\ee
It is sufficient to find an appropriate representation at the
one-oscillator level. Indeed, assume we have found ${U}_{\Lambda,y}$
satisfying 
\be
e^{\pm 2i\Theta(\Lambda,\bbox k)}e^{ik\cdot y}
a(\bbox {\Lambda^{-1}k},\pm)
=
{U}_{\Lambda,y}^{\dag} a(\bbox {k},\pm){U}_{\Lambda,y}.
\ee
Then 
\be
{\uu U}_{\Lambda,y}
&=&
\bigoplus_{N=1}^\infty
\underbrace{U_{\Lambda,y}\otimes\dots\otimes U_{\Lambda,y}}_N.
\ee

\subsection{Four-translations}

The definition of four momentum for a single harmonic oscillator is 
\be
P_a
&=&
\int d\Gamma(\bbox k)k_a|\bbox k\rangle\langle \bbox k|\otimes h
\ee
where 
\be
h=\frac{1}{2}\sum_s\Big(a^{\dag}_sa_s+a_s\,a^{\dag}_s\Big)
=\sum_s h_s.
\ee
One immediately verifies that 
\be
e^{iP\cdot x}a(\bbox k,s)e^{-iP\cdot x}
&=&
a(\bbox k,s) e^{-ix\cdot k}\\
e^{iP\cdot x}a(\bbox k,s)^{\dag}e^{-iP\cdot x}
&=&
a(\bbox k,s)^{\dag} e^{ix\cdot k}
\ee
implying 
\be
U_{\bbox 1,y}=e^{iy\cdot P}.
\ee
Consequently, the generator of four-translations corresponding to 
$
\uu U_{\bbox 1,y}=e^{iy\cdot \uu P}
$
is 
$
\uu P_a=\oplus_1 P_a
$ 
and 
\be
e^{i{\uu P}\cdot x}\uu a(\bbox k,s)^{\dag}e^{-i{\uu P}\cdot x}
&=&
\uu a(\bbox k,s)^{\dag} e^{ix\cdot k}\\
e^{i{\uu P}\cdot x}\uu a(\bbox k,s)e^{-i{\uu P}\cdot x}
&=&
\uu a(\bbox k,s) e^{-ix\cdot k}.
\ee
The $x$-dependence of field operators can be introduced via $\uu P$:
\be
{\uu F}_{ab}(x)
&=&
e^{i{\uu P}\cdot x}
{\uu F}_{ab}
e^{-i{\uu P}\cdot x}
\ee
\subsection{Rotations and boosts}

To find an analogous representation of 
\be
\uu a(\bbox {k},\pm)
\mapsto
e^{\pm 2i\Theta(\Lambda,\bbox k)}
\uu a(\bbox {\Lambda^{-1}k},\pm)
=
{\uu U}_{\Lambda,0}^{\dag} \uu a(\bbox {k},\pm){\uu U}_{\Lambda,0}
\ee
we define
\be
U_{\Lambda,0}
&=&
\exp\Big(\sum_s 2is\int d\Gamma(\bbox k)\Theta(\Lambda,\bbox k)|\bbox
k\rangle \langle \bbox k|\otimes h_s\Big)
\Big(\sum_r\int d\Gamma(\bbox p)|\bbox p,r\rangle
\langle \bbox {\Lambda^{-1}p},r|\otimes  1\Big).
\ee
Finally the transformations of the field tensor are 
\be
{\uu U}_{\Lambda,0}^{\dag}{\uu F}_{ab}(x){\uu U}_{\Lambda,0}
&=&
\Lambda{_a}{^c}\Lambda{_b}{^d}{\uu F}_{cd}(\Lambda^{-1}x)\\
{\uu U}_{\bbox 1,y}^{\dag}{\uu F}_{ab}(x){\uu U}_{\bbox 1,y}
&=&
{\uu F}_{ab}(x-y)
\ee
The zero-energy part of $\uu P$ can be removed by a unitary
transformation leading to a {\it vacuum picture\/} dynamics (cf.
\cite{II}). We will describe this in more detail after having
discussed the properties of non-canonical states. 

\section{States and their Poincar\'e transformations}

It is clear that in order to control transformation properties of
states it is sufficient to discuss single-oscillator representations.
We shall start with single-oscillator states and then extend them to
many oscillators.  

\subsection{Representation in the one-oscillator sector}

The one-oscillator Hilbert space consists of functions $f$ satisfying 
\be
\sum_{n_+,n_-=0}^\infty\int d\Gamma(\bbox k)|f(\bbox k,n_+,n_-)|^2
<\infty.
\ee
We will write them in the Dirac notation as  
\be
|f\rangle
&=&
\sum_{n_\pm}\int d\Gamma(\bbox k)f(\bbox k,n_+,n_-)|\bbox k,n_+,n_-\rangle.
\ee
The representation of the Poincar\'e group is
\be
|f\rangle
\mapsto
U_{\Lambda,y}|f\rangle
&=&
U_{\bbox 1,y}U_{\Lambda,0}|f\rangle\nonumber\\
&=&
\sum_{n_\pm}\int d\Gamma(\bbox k)f(\bbox {\Lambda^{-1}k},n_+,n_-)
e^{2i(n_+-n_-)\Theta(\Lambda,\bbox {k})}e^{ik\cdot
y(n_++n_-+1/2)} |\bbox {k},n_+,n_-\rangle. 
\ee
The latter formula can be written as 
\be
f(\bbox k,n_+,n_-)
\mapsto
U_{\Lambda,y}f(\bbox k,n_+,n_-)
&=&
e^{ik\cdot y(n_++n_-+1/2)}
e^{2i(n_+-n_-)\Theta(\Lambda,\bbox {k})}
f(\bbox {\Lambda^{-1}k},n_+,n_-)\label{U1}
\ee
or 
\be
U_{\Lambda,y}|f\rangle
=
|U_{\Lambda,y}f\rangle.\label{U2}
\ee
The form (\ref{U1}) is very similar to the zero-mass spin-1
representation (\ref{unit}), the difference being in the multiplier
$n_++n_-+1/2$. One can check by a straightforward calculation that
(\ref{U1}) defines a representation of the group.  

\subsection{Generators and vacuum picture}

Denote by $K_a$ and $L_{ab}+S_{ab}$ the generators of 4-translations
and $SL(2,C)$ of the standard zero-mass spin-1 unitary representation
of the Poincar\'e group. $L_{ab}$ denotes the orbital part of the
generator. The generators of (\ref{U1}) are then
\be
P^a&=&K_a\otimes h\label{gen1}\\
J^{ab}&=&L^{ab}\otimes 1+ S{^{ab}}{_{ss}}\otimes h_s.\label{gen2}
\ee
$S{^{ab}}{_{ss'}}$ are matrix elements of $S{^{ab}}$ (which is a
diagonal $p$-dependent matrix).
Denote by $\tilde S_{ab}$ the generators of the $(1/2,1/2)$ spinor
representation of $SL(2,C)$, i.e.
$\Lambda=\exp\big(i\xi^{ab}\tilde S_{ab}/2\big)$. The generators
of the unitary representation are defined by 
\be
P_af(\bbox k,n_+,n_-)
&=&
-i\frac{\partial}{\partial y^a}
U_{\Lambda,y}f(\bbox k,n_+,n_-)\big|_{\xi,y=0}\\
J_{ab}f(\bbox k,n_+,n_-)
&=&
-i\frac{\partial}{\partial \xi^{ab}}
U_{\Lambda,y}f(\bbox k,n_+,n_-)\big|_{\xi,y=0}
\ee
In what follows we will work in a ``vacuum picture", i.e with unitary
transformations 
\be
f(\bbox k,n_+,n_-)
\mapsto
V_{\Lambda,y}f(\bbox k,n_+,n_-)
&=&
e^{i(n_++n_-)k\cdot y}
e^{2i(n_+-n_-)\Theta(\Lambda,\bbox {k})}
f(\bbox {\Lambda^{-1}k},n_+,n_-)\label{Uvac}.
\ee
The transition 
\be
U_{\Lambda,y}\mapsto V_{\Lambda,y}=
W_{y}^{\dag}U_{\Lambda,y}
\ee
is performed by means of the unitary transformation 
which commutes with non-CCR creation and annihilation operators.  

Let us stress that the fact that we ``remove" the zero-energy parts
from generators does not mean that energy of vacuum is zero. The
vacuum picture is in a sense a choice of representation co-moving
with vacuum.  

\subsection{Vacuum states}

Vacuum states are all the states which are annihilated by all
annihilation operators. At the one-oscillator level these are states
of the form 
\be
|O\rangle=\int d\Gamma(\bbox k)O(\bbox k)|\bbox k,0,0\rangle.
\ee
Even in the vacuum picture the vacuum states are not Poincar\'e
invariant since 
\be
V_{\Lambda,y}O(\bbox k)
&=&
O(\bbox {\Lambda^{-1}k})
\ee
which means they transform as a 4-translation-invariant
scalar field. We will often meet the expression 
$Z(\bbox k)=|O(\bbox k)|^2$ describing the probability density of the
``zero modes".  

\subsection{Coherent states}

An analogue of the standard coherent (or ``semiclassical") state is 
at the 1-oscillator level 
\be
|O_\alpha\rangle
&=&
\int d\Gamma(\bbox k)O(\bbox {k})
|\bbox {k}\rangle|\alpha (\bbox {k},+),\alpha (\bbox {k},-)\rangle
\ee
where 
\be
a_s|\alpha (\bbox {k},+),\alpha (\bbox {k},-)\rangle
=
\alpha (\bbox {k},s)|\alpha (\bbox {k},+),\alpha (\bbox {k},-)\rangle 
\ee
Its explicit form in the basis of eigenstates of the oscillator is 
\be
|O_\alpha\rangle
&=&
\int d\Gamma(\bbox k)
O(\bbox k)\sum_{n_+,n_-=0}^\infty\frac{\alpha(\bbox k,+)^{n_+}}{\sqrt{n_+!}}
e^{-|\alpha(\bbox k,+)|^2/2}
\frac{\alpha(\bbox k,-)^{n_-}}{\sqrt{n_-!}}
e^{-|\alpha(\bbox k,-)|^2/2}
|\bbox k\rangle|n_+\rangle|n_-\rangle\\
&=&
\sum_{n_+,n_-}\int d\Gamma(\bbox k)
O_\alpha(\bbox k,n_+,n_-)
|\bbox k,n_+,n_-\rangle
\ee
where 
\be
O_\alpha(\bbox k,n_+,n_-)
=
\frac{1}{\sqrt{n_+!n_-!}}O(\bbox k)
\alpha(\bbox k,+)^{n_+}\alpha(\bbox k,-)^{n_-}
e^{-\sum_\pm|\alpha(\bbox k,\pm)|^2/2}
\ee
The average of the 1-oscillator field operator evaluated in such a
coherent state is 
\be
\langle O_\alpha|{^-}\hat F_{ab}(x)|O_\alpha\rangle
&=&
\int d\Gamma(\bbox k)e_{ab}(\bbox k)
Z(\bbox k)
\Big(\alpha(\bbox k,-)e^{-ik\cdot x}
+
\overline{\alpha(\bbox k,+)}e^{ik\cdot x}\Big).
\ee
The Poincar\'e transformation of the state implies 
\be
\langle O_\alpha|V_{\Lambda,y}^{\dag}
{^-}\hat F_{ab}(x)V_{\Lambda,y}|O_\alpha\rangle
=
\Lambda{_a}{^c}\Lambda{_b}{^d}
\langle O_\alpha|{^-}\hat F_{cd}\big(\Lambda^{-1}(x-y)\big)|O_\alpha\rangle
\ee
The coherent-state wave function transforms by 
\be
V_{\Lambda,y}O_\alpha(\bbox k,n_+,n_-)
&=&
e^{i(n_++n_-)k\cdot y}
e^{2i(n_+-n_-)\Theta(\Lambda,\bbox {k})}
O_\alpha(\bbox {\Lambda^{-1}k},n_+,n_-)\\
&=&
O(\bbox {\Lambda^{-1}k})
\prod_{s=\pm}
e^{in_sk\cdot y}
e^{2isn_s\Theta(\Lambda,\bbox {k})}
\frac{1}{\sqrt{n_s!}}
\alpha(\bbox {\Lambda^{-1}k},s)^{n_s}
e^{-|\alpha(\bbox {\Lambda^{-1}k},s)|^2/2}
\\
&=&
O(\bbox {\Lambda^{-1}k})
\prod_{s=\pm}
\frac{1}{\sqrt{n_s!}}\Big(
T_{\Lambda,y}\alpha(\bbox {k},s)\Big)^{n_s}
e^{-|T_{\Lambda,y}\alpha(\bbox {k},s)|^2/2}
\ee
where 
\be
\alpha(\bbox {k},s)\mapsto T_{\Lambda,y}\alpha(\bbox {k},s)
\ee
is the spin-1 massless unitary representation (\ref{unit}).
Using this result we get again 
\be
\langle O_\alpha|V_{\Lambda,y}^{\dag}
{^-}\hat F_{ab}(x)V_{\Lambda,y}|O_\alpha\rangle
&=&
\int d\Gamma(\bbox k)e_{ab}(\bbox k)
Z(\bbox {\Lambda^{-1}k})
\Big(T_{\Lambda,y}\alpha(\bbox {k},-)
e^{-ik\cdot x}
+
\overline{T_{\Lambda,y}\alpha(\bbox {k},+)}
e^{ik\cdot x}\Big)\\
&=&
\Lambda{_a}{^c}\Lambda{_b}{^d}
\langle O_\alpha|{^-}\hat F_{cd}\big(\Lambda^{-1}(x-y)\big)|O_\alpha\rangle
\ee
showing that those somewhat counter-intuitive forms of
$U_{\Lambda,y}$ and $V_{\Lambda,y}$ are consistent with passive
$T_{\Lambda,y}$ transformations of classical wave functions.  

With $a(\beta)$ and $a(\beta)^{\dag}$ given by
(\ref{a(f)})-(\ref{a(f)^+}) we define the displacement operator 
\be
{\cal D}(\beta)
&=&
e^{a(\beta)^{\dag}-a(\beta)}\\
&=&
\exp\Big(\sum_s\int d\Gamma(\bbox k)
\big(\beta(\bbox k,s) a(\bbox k,s)^{\dag}
-\overline{\beta(\bbox k,s)} a(\bbox k,s)\big)\Big)\\
&=&
\int d\Gamma(\bbox k)
|\bbox k,s\rangle\langle\bbox k,s|\otimes
e^{\sum_s\big(\beta(\bbox k,s) a^{\dag}_s
-
\overline{\beta(\bbox k,s)} a_s\big)}
\ee
which performs a shift of the classical wave function
\be
{\cal D}(\beta)|O_\alpha\rangle=|O_{\alpha+\beta}\rangle
\ee
and commutes with $I_{\bbox k}$:
\be
{\cal D}(\beta)^{\dag}I_{\bbox k} {\cal D}(\beta)=I_{\bbox k}.
\ee
Vacuum states are also coherent states corresponding to $\alpha=0$.

\subsection{Multi-oscillator coherent states}

Consider a family $\alpha_N(\bbox {k},s)$, $N=1,2,\dots$ of functions and the 
state 
\be
|\uu O_\alpha\rangle
&=&
\bigoplus_{N=1}^\infty
\sqrt{p_N}\underbrace{|O_{\alpha_N}\rangle\otimes\dots
\otimes |O_{\alpha_N}\rangle}_N\label{mocs}
\ee
where 
\be
|O_{\alpha_N}\rangle
&=&
\int d\Gamma(\bbox k)O(\bbox {k})
|\bbox {k}\rangle|\alpha_N (\bbox {k},+),\alpha_N (\bbox {k},-)\rangle
\ee
and $\sum_{N=1}^\infty p_N=1$. 
Taking, for example, $\alpha_N(\bbox {k},s)=\alpha(\bbox {k},s)/\sqrt{N}$ 
we find 
\be
\langle\uu O_\alpha|{^-}{\uu {\hat F}}_{ab}(x)|\uu O_\alpha\rangle
&=&
\int d\Gamma(\bbox k)e_{ab}(\bbox k)
Z(\bbox k)
\Big(\alpha(\bbox k,-)e^{-ik\cdot x}
+
\overline{\alpha(\bbox k,+)}e^{ik\cdot x}\Big)
=
\langle O_\alpha|{^-}\hat F_{ab}(x)|O_\alpha\rangle
\ee
i.e. the same result as in the 1-oscillator case. 

A multi-oscillator displacement operator is 
\be
{\uu {\cal D}}(\beta)
&=&
\bigoplus_{N=1}^\infty \underbrace{{\cal D}(\beta_N)\otimes
\dots\otimes {\cal D}(\beta_N)}_N
=
e^{\uu a(\beta)^{\dag}-\uu a(\beta)},\label{uu D}
\ee
$\beta_N(\bbox {k},s)=\beta(\bbox {k},s)/\sqrt{N}$,
implying  
\be
\uu {\cal D}(\beta)|\uu O_\alpha\rangle
&=&
|\uu O_{\alpha+\beta}\rangle\\
{\uu {\cal D}}(\beta)^{\dag}\uu a(\bbox p,s){\uu {\cal D}}(\beta)
&=&
\uu a(\bbox p,s)+\beta(\bbox p,s)\uu I_{\bbox p}\\
{\uu {\cal D}}(\beta)^{\dag}\uu I_{\bbox p}{\uu {\cal D}}(\beta)
&=&
\uu I_{\bbox p}.
\ee
The fact that $\alpha_N(\bbox {k},s)=\alpha(\bbox {k},s)/\sqrt{N}$
will be shown to be of crucial importance for the question of
statistics of excitations of multi-oscillator coherent states. Let us
note that a similar property of coherent states was found in \cite{I}
when we employed the definition in terms of eigenstates of
annihilation operators.  

\subsection{Multi-oscillator vacua}

Vacuum consists of states with $n_\pm=0$, i.e. 
with all the oscillators in their ground states. 
Of particular interest, due to its simplicity, is the following vacuum state
\be
|\uu O\rangle
&=&
\bigoplus_{N=1}^\infty
\sqrt{p_N}\underbrace{|O\rangle\otimes\dots
\otimes |O\rangle}_N\label{uu O}
\ee
where 
\be
|O\rangle
&=&
\int d\Gamma(\bbox k)O(\bbox {k})
|\bbox {k},0,0\rangle
\ee
Such a vacuum is simultaneously a particular case of a coherent state
with $\alpha(\bbox k,s)=0$. Coherent states are related to the vacuum
state via the displacement operator 
\be
{\uu {\cal D}}(\alpha)|\uu O\rangle
&=&
\bigoplus_{N=1}^\infty
\sqrt{p_N}\underbrace{|O_{\alpha_N}\rangle\otimes\dots
\otimes |O_{\alpha_N}\rangle}_N\\
&=&
|\uu O_\alpha\rangle.
\ee

\subsection{Normalized 1-photon states}

Consider the vector 
\be
\uu a(f)^{\dag}|\uu O\rangle.
\ee
Choosing the particular form (\ref{uu O}) 
we find 
\be
\langle \uu O|\uu a(f)\uu a(g)^{\dag}|\uu O\rangle
&=&
\sum_{s}\int d\Gamma(\bbox k) Z(\bbox k)
\overline{f(\bbox k,s)}g(\bbox k,s)
\\
&=&
\langle fO|gO\rangle=:\langle f|g\rangle_Z.
\ee
$fO$ denotes the pointlike product 
$fO(\bbox k,s)=O(\bbox k)f(\bbox k,s)$. Since anyway only the modulus
$|O(\bbox k)|=Z(\bbox k)^{\frac{1}{2}}$ 
occurs in the above scalar products one can 
also work with $f_B(\bbox k,s)=Z(\bbox k)^{\frac{1}{2}}f(\bbox k,s)$. 
The relation between $f_B$ and $f$ resembles the one between the bare
and renormalized fields \cite{Weinberg}. 
We believe this is more than just an analogy. 

Thinking of bases in the Hilbert space 
one can take functions $f_i$ satisfying 
\be
\langle f_i|f_j\rangle_Z=\delta_{ij}\label{delta}=
\langle f_{Bi}|f_{Bj}\rangle.
\ee
\subsection{Normalization of multi-photon states}

Normalization of multi-photon states is more complicated. In this
section we will discuss this point in detail since the argument we
give is very characteristic for the non-canonical framework. It can
be used to show that in the thermodynamic limit of a large number of
oscillators the non-CCR perturbation theory tends to the CCR one but
in a version which is automatically regularized \cite{II}. We will
also use a similar trick to show that the multi-oscillator coherent
states have, again in the thermodynamic limit, Poissonian statistics of
excitations.  

Denote by $\sum_\sigma$ the sum over all the permutations of the set
$\{1,\dots,m\}$.
\medskip

\noindent
{\bf Theorem 1.} Consider the vacuum state (\ref{uu O}) with $p_N=1$
for some $N$. Then
\be
{}&{}&\lim_{N\to\infty}
\langle \uu O|\uu a(f_1)\dots \uu a(f_m)\uu a(g_1)^{\dag}\dots
\uu a(g_m)^{\dag}|\uu O\rangle
=
\sum_{\sigma}
\langle f_1|g_{\sigma(1)}\rangle_Z
\dots 
\langle f_m|g_{\sigma(m)}\rangle_Z\nonumber\\
&{}&\pp{==}=
\sum_{\sigma}\sum_{s_1\dots s_m}\int d\Gamma(\bbox k_1)Z(\bbox k_1)\dots
d\Gamma(\bbox k_m)Z(\bbox k_m)
\overline{f_1(\bbox k_1,s_1)}\dots 
\overline{f_m(\bbox k_m,s_m)}
g_{\sigma(1)}(\bbox k_1,s_1)
\dots 
g_{\sigma(m)}(\bbox k_m,s_m)\nonumber
\ee
{\it Proof:\/}  
The scalar product of two general unnormalized multi-photon states
is
\be
{}&{}&\langle \uu O|\uu a(f_1)\dots \uu a(f_m)\uu a(g_1)^{\dag}\dots
\uu a(g_m)^{\dag}|\uu O\rangle\nonumber\\ 
&{}&\pp =
=
\sum_{\sigma}\sum_{s_1\dots s_m}\int d\Gamma(\bbox k_1)\dots
d\Gamma(\bbox k_m)\overline{f_1(\bbox k_1,s_1)}\dots 
\overline{f_m(\bbox k_m,s_m)}
g_{\sigma(1)}(\bbox k_1,s_1)
\dots 
g_{\sigma(m)}(\bbox k_m,s_m)
\langle \uu O|
\uu I_{\bbox k_1}\dots \uu I_{\bbox k_m}
|\uu O\rangle\nonumber\\
&{}&\pp =
=
\sum_{\sigma}\sum_{s_1\dots s_m}\int d\Gamma(\bbox k_1)\dots
d\Gamma(\bbox k_m)\overline{f_1(\bbox k_1,s_1)}\dots 
\overline{f_m(\bbox k_m,s_m)}
g_{\sigma(1)}(\bbox k_1,s_1)
\dots 
g_{\sigma(m)}(\bbox k_m,s_m)
\nonumber\\
&{}&
\pp =
\times
\frac{1}{N^m}
\underbrace{
\langle O|\dots\langle O|}_N
\Big(I_{\bbox k_1}\otimes \dots\otimes I+
\dots +I\otimes\dots\otimes I_{\bbox k_1}\Big)
\dots
\Big(I_{\bbox k_m}\otimes \dots\otimes I+
\dots +I\otimes\dots\otimes I_{\bbox k_m}\Big)
\underbrace{
|O\rangle\dots |O\rangle}_N\label{multi-norm}
\ee
Further analysis of (\ref{multi-norm}) can be simplified by the
following notation: 
\be
1_{k_j} &=&  I_{\bbox k_j}\otimes \dots\otimes I\nonumber
\\
2_{k_j} &=&  I\otimes I_{\bbox k_j}\otimes \dots\otimes I\nonumber
\\
        &\vdots& \nonumber\\
N_{k_j} &=& I\otimes\dots\otimes I_{\bbox k_j}\nonumber
\ee
with $j=1,\dots,m$; the sums-integrals $\sum_{s_j}\int
d\Gamma(\bbox k_j)$ are denoted by $\sum_{k_j}$.  Then
(\ref{multi-norm}) can be written as
\be
&{}&
\sum_\sigma
\sum_{k_1\dots k_m}
\overline{f_1(k_1)}\dots 
\overline{f_m(k_m)}
g_{\sigma(1)}(k_1)
\dots 
g_{\sigma(m)}(k_m)
\frac{1}{N^m}
\sum_{A\dots Z=1}^N
\underbrace{
\langle O|\dots\langle O|}_N
A_{k_1}\dots Z_{k_m}
\underbrace{
|O\rangle\dots |O\rangle}_N\label{az}
\ee
Since $m$ is fixed and we are interested in the limit $N\to\infty$ we
can assume that $N>m$. Each element of the sum over 
$A_{k_1}\dots Z_{k_m}$ in (\ref{az}) can be associated with a unique
point $(A,\dots,Z)$ in an $m$-dimensional lattice embedded in a cube
with edges of length $N$.  

Of particular interest are those points of the cube, the
coordinates of which are all different. Let us denote the subset of
such points by 
$C_0$. For $(A,\dots,Z)\in C_0$ 
\be
\underbrace{
\langle O|\dots\langle O|}_N
A_{k_1}\dots Z_{k_m}
\underbrace{
|O\rangle\dots |O\rangle}_N
=
Z(\bbox k_1)\dots Z(\bbox k_m)\label{contr}
\ee
no matter what $N$ one considers and what are the numerical
components in $(A,\dots,Z)$. (This makes sense only for $N>m$; otherwise 
$C_0$ would be empty). Therefore each element of $C_0$ produces an
identical contribution (\ref{contr}) to (\ref{az}). Let us denote the
number of points in $C_0$ by $N_0$. 

The sum (\ref{az}) can be now written as 
\be
{}&{}&
\sum_\sigma
\sum_{k_1\dots k_m}
\overline{f_1(k_1)}\dots 
\overline{f_m(k_m)}
g_{\sigma(1)}(k_1)
\dots 
g_{\sigma(m)}(k_m)
{\cal P}_0
Z(\bbox k_1)\dots Z(\bbox k_m)\nonumber\\
&{}&\pp{==}
+
\sum_\sigma
\sum_{k_1\dots k_m}
\overline{f_1(k_1)}\dots 
\overline{f_m(k_m)}
g_{\sigma(1)}(k_1)
\dots 
g_{\sigma(m)}(k_m)
\frac{1}{N^m}
\sum_{(A\dots Z)\notin C_0}
\underbrace{
\langle O|\dots\langle O|}_N
A_{k_1}\dots Z_{k_m}
\underbrace{
|O\rangle\dots |O\rangle}_N.\label{az'}
\ee
The coefficient ${\cal P}_0=\frac{N_0}{N^m}$ represents a probability
of $C_0$ in the cube. 
The elements of the remaining sum over $(A\dots Z)\notin C_0$ can be also
grouped into classes according to the values of 
$
\langle O|\dots\langle O|
A_{k_1}\dots Z_{k_m}
|O\rangle\dots |O\rangle$. There are $m-1$ such different classes,
each class has its associated probability ${\cal P}_j$, $0<j\leq
m-1$, which will appear  in the sum in an analogous role as 
${\cal P}_0$. 

The proof is completed by the observation that 
\be
\lim_{N\to\infty}{\cal P}_0 &=& 1,\\
\lim_{N\to\infty}{\cal P}_j &=& 0,\quad 0<j.
\ee
Indeed, the probabilities are unchanged if one rescales the cube to 
$[0,1]^m$. The probabilities are computed by means of an
$m$-dimensional uniformly distributed measure. 
$N\to\infty$ corresponds to the continuum
limit, and in this limit the sets of points of which at least two
coordinates are equal are of $m$-dimensional measure zero. 
\rule{5pt}{5pt}
\medskip

\noindent
{\it Comments:\/} (a) The thermodynamic limit is naturally equipped
with the scalar product yielding orthogonality relation of the form
(\ref{delta}). However, for small $N$ there will be differences if $m$
is large. On the other hand if $N$ is sufficiently large then the values
of $m$ for which the corrections are non-negligible must be also
large. But then a classical limit will be justified and the use of
non-canonical coherent states should again give the correct
description. (b) Concrete values of ${\cal P}_j$ for some small $m$
were given in \cite{I}.
For $m=2$: ${\cal P}_0=1-1/N$,  ${\cal P}_1=1/N$; 
for $m=3$: ${\cal P}_0=1-3/N+2/N^2$,  ${\cal P}_1=3/N-3/N^2$, 
${\cal P}_2=1/N^2$. In general we do not have to assume that $p_N=1$.
If $p_N$ are general probabilities then the coefficients involve
averages. For $m=2$: ${\cal P}_0=1-\langle 1/N\rangle$, ${\cal
P}_1=\langle 1/N\rangle$; 
for $m=3$: ${\cal P}_0=1-\langle 3/N\rangle+\langle 2/N^2\rangle$,  
${\cal P}_1=\langle 3/N\rangle-\langle 3/N^2\rangle$, 
${\cal P}_2=\langle 1/N^2\rangle$, where $\langle 1/N\rangle=\sum_N p_N/N$
etc. The normalization in terms of $\langle\cdot|\cdot\rangle_Z$ is
then obtained under the assumption that all those averages vanish,
which can hold only approximately, meaning that the probability $p_N$
is peaked in a region of large $N$s. 

%aaaaaaaaa
\subsection{Space of states as a vector bundle}

An analysis of perturbation theory given in \cite{II} suggested that
non-canonical probabilities have to be computed as if the space of
vacua was one dimensional (a ``unique vacuum"). In the non-canonical
theory it is essential
that vacuum is represented by an infinite dimensional subspace of all
the states annihilated by all annihilation operators. Otherwise it is
impossible to associate the entire spectrum of frequencies with a single
oscillator. 

The way one removes the zero-energy part from the four-momentum 
shows that the role of a reference state is a ``moving
vacuum" whose dynamics is given by the zero-enegy part of
four-momentum (the vacuum picture). 

The construction of multi-photon states shows that, having a vacuum
state $|\uu O\rangle$ 
representing an ensemble of ground-state oscillators, one can
introduce a Fock-type structure representing excitations of the
ensemble. Obviously, the Fock spaces of multiphoton states are
different for different vacua. 

All of this suggests a vector-bundle structure where the set of
vacuum states is a base space and Fock spaces are fibers. The
probabilities are calculated in fibers. The Poincar\'e group acts in
the entire bundle. The zero-energy part of four-momenta generates a
motion in the base space and plays essentially a role of a bundle connection.

\section{Statistics of excitations}

It is an experimental fact that laser beams produce Poissonian
statistics of photocounts. At the theoretical level of canonical
quantum optics the Poisson 
distribution follows trivially from the form of canonical coherent
states. In the non-canonical case the exact Poisson statistics is
characteristic of the single-oscillator ($N=1$) sector. For $1<N<\infty$ the
statistics of excitations is non-Poissonian. At the other extreme 
is the thermodynamic limit for multi-oscillator states. 
In what follows we will show that in
the limit $N\to\infty$ one recovers {\it the same\/} Poisson distribution as
for $N=1$. This, at a first glance unexpected, result justifying our
definitions in terms of displacement operators is a consequence of
certain classical universality properties of the Poisson distribution.

We will also
return to the question of thermal states and the Planck formula. In
\cite{I} it was argued that non-CCR quantization implies deviations
from the black-body law. However, a consistent interepretation of the
field in terms of the thermodynamic limit shows that no deviations
should be expected. 

\subsection{Multi-oscillator coherent states}

To study the thermodynamic limit of multi-oscillator coherent states
we simplify the discussion by taking an exactly $N$-oscillator
coherent state (\ref{mocs}) ($N\gg 1$ is fixed and $p_N=1$), i.e. 
\be
|\uu O_\alpha\rangle
&=&
\underbrace{|O_{\alpha_N}\rangle\otimes\dots
\otimes |O_{\alpha_N}\rangle}_N
\ee
where 
\be
|O_{\alpha_N}\rangle
&=&
\int d\Gamma(\bbox k)O(\bbox {k})
|\bbox {k},s\rangle|
\alpha (\bbox {k},+)/\sqrt{N}\rangle|\alpha (\bbox {k},-)/\sqrt{N}\rangle.
\ee
The average number of excitations in this state is 
\be
\langle n\rangle
&=&
\sum_{s}\int d\Gamma(\bbox k)Z(\bbox {k})
|\alpha (\bbox {k},s)|^2.
\ee
The simplest case is the one where $\alpha (\bbox {k},s)=\alpha={\rm
const}$. Then $\langle n\rangle=|\alpha|^2$ and the statistics of
excitations of single-oscillator coherent states $|O_{\alpha_N}\rangle$
is Poissonian with the distribution
$p_n=e^{-|\alpha_N|^2}|\alpha_N|^{2n}/n!$ 

$m$ excitations distributed in the ensemble of $N$ oscillators can be
represented by the ordered $m$-tuple $(j_1,\dots,j_m)$, 
$1\le j_1\le\dots\le j_m \le N$.  For
example, for $m=10$, $N=12$, the point $(2,2,2,5,5,7,7,7,11,11)$
represents 10 excitations distributed in the
ensemble of 12 oscillators as follows: 3 excitations in 2nd
oscillator, 2 in the 5th one, 3 in the 7th, and 2 in the 11th. Such
points form a subset of the cube $[0,N]^m$, the interior of the set
corresponding to points whose all the indices are different. The latter
means that the interior represents situations where there are $m$
oscillators excited, and each of them is in the first excited state.
The boundary of this set consists of points representing at least one
oscillator in a higher excited state. 
Probabilities of events represented by points with 
the same numbers of repeated indices must be identical due to
symmetries. 
Intuitively, the Poissonian
statistics of the thermodynamic limit follows 
from the fact that the probability of finding a point belonging to
the boundary tends to zero as $N$ increases. The statistics is
dominated by Bernoulli-type processes with probabilities
related to the two lowest energy levels of a single oscillator in a
coherent state.

To make the argument more formal we introduce the following notation:
\be
X_m^{(N)}
&=&
\{x\in {\bbox N}^m; m\geq 1,x=(j_1,\dots,j_m), 1\le j_1\le\dots\le j_m \le
N\}
\nonumber\\
X_{n_1\dots n_k}^{(N)}
&=&
\{x\in X_m^{(N)}; 
x=(\underbrace{i_1,\dots,i_1}_{n_1}, 
\dots
\underbrace{i_k,\dots,i_k}_{n_k}),
i_1<\dots <i_k
\}
\nonumber\\
Y_m^{(N)}
&=&
\bigcup_{(n_1\dots n_k)\neq (1\dots 1)}X_{n_1\dots n_k}^{(N)}
\ee
If we add a single-element set $X_0^{(N)}$ containing the event
representing $N$ oscillators in their ground states we can represent
the set of all the events by the disjoint sum 
\be
X^{(N)}
&=&
\bigcup_{m=0}^\infty X_m^{(N)}
\ee
The probability of finding the partition $m=n_1+\dots +n_k$ is 
\be
P(X_{n_1\dots n_k}^{(N)})&=&
N_{n_1\dots n_k}p_{n_1}\dots p_{n_k}p_0^{N-k}
\nonumber\\
&=&
N_{n_1\dots n_k}\frac{e^{-N|\alpha_N|^2}|\alpha_N|^{2m}}{n_1!\dots n_k!}
\ee
where $N_{n_1\dots n_k}$ is the number of elements of $X_{n_1\dots
n_k}^{(N)}\subset X_{m}^{(N)}$.  The probability
that $m$ excitations are found is 
\be
P(X_{m}^{(N)})&=&\sum_{n_1\dots n_k} P(X_{n_1\dots n_k}^{(N)}),
\ee
the sum being over all the partitions of $m$. Denote by 
$
P(Y_{m}^{(N)}|X_{m}^{(N)})
$
the conditional probability of finding at least one oscillator in the
2nd or higher excited state under the condition that the sum of excitations
is $m>1$. We first prove the following 

\medskip
\noindent 
{\bf Lemma 1.} 
\be
\lim_{N\to\infty} P(Y_{m}^{(N)}|X_{m}^{(N)})=0.
\ee
{\it Proof\/}: Since $Y_{m}^{(N)}\cap X_{m}^{(N)}=Y_{m}^{(N)}$ one
finds 
\be
P(Y_{m}^{(N)}|X_{m}^{(N)})
&=&
\frac{\sum_{(n_1\dots n_k)\neq (1\dots 1)} P(X_{n_1\dots n_k}^{(N)})}
{\sum_{n_1\dots n_k} P(X_{n_1\dots n_k}^{(N)})}
\nonumber\\
&=&
\Big[1+N_{1\dots 1}\Big(
\sum_{(n_1\dots n_k)\neq (1\dots 1)} \frac{N_{n_1\dots n_k}}
{n_1!\dots n_k!}\Big)^{-1}\Big]^{-1}
\nonumber\\
&<&
\Big[1+N_{1\dots 1}\Big(
\sum_{(n_1\dots n_k)\neq (1\dots 1)} N_{n_1\dots n_k}
\Big)^{-1}\Big]^{-1}
\ee
However, 
\be
\lim_{N\to\infty}
\frac{\sum_{(n_1\dots n_k)\neq (1\dots 1)} N_{n_1\dots n_k}}
{N_{1\dots 1}}=0
\ee
on the basis of the geometric argument we gave in the previous
section. This ends the proof. \rule{5pt}{5pt}

The main result of this section is the following version of the well
known Poisson theorem:

\medskip
\noindent 
{\bf Theorem 2.} 
Assume that $\alpha(\bbox k,s)=\alpha={\rm const}$. Then 
\be
\lim_{N\to\infty} P(X_{m}^{(N)})
=
\frac{e^{-|\alpha|^2}|\alpha|^{2m}}{m!}. 
\ee
{\it Proof\/}: 
As an immediate consequence of the lemma we find
\be
\lim_{N\to\infty} P(X_{m}^{(N)})&=&\lim_{N\to\infty} 
P(X_{m}^{(N)}-Y_{m}^{(N)})
\ee
which means that in the thermodynamic limit we can treat excitations of the
oscillators to 2nd and higher excited levels as events whose
probability is zero. The probabilities of ground and first excited states
follow from the single-oscillator Poisson distributions but
conditioned by the fact that only the lowest two levels are taken
into account.  

We thus arrive at the standard Poisson process with 
\be
P_N &=& \frac{p_1}{p_0+p_1}=\frac{|\alpha_N|^2}{1+|\alpha_N|^2}
=
\frac{|\alpha|^2/N}{1+|\alpha|^2/N}
\ee
and $\lim_{N\to\infty}NP_N=|\alpha|^2$. \rule{5pt}{5pt}

\subsection{Thermal states}

A single-oscillator free-field Hamiltonian $H$ has the usual eigenvalues 
\be
E(\omega,n)=\hbar\omega \Big(n+\frac{1}{2}\Big).
\ee
The eigenvalues of the free-field Hamiltonian $\uu H$ at the
multi-oscillator level are sums of the 
single-oscillator ones. In \cite{I} it was assumed that the
Boltzmann-Gibbs distribution of thermal radiation should be
constructed in terms of $\uu H$. Let us note, however, that such a
construction makes use of $\uu H$ as if it was a Hamiltonian of a
single element of a statistical ensemble. The discussion of the
thermodynamic limit we have given above, as well as the results of
\cite{II}, suggest that $\uu H$ is the Hamiltonian of the entire
ensemble of systems described by $H$, and it is $H$ and not $\uu H$
which should be used in the Boltzmann-Gibbs distribution. Then, of
course, the result will be the standard one and no deviations from
the Planck formula will occur. 

\section{Commutators of 4-potentials and locality}

A straightforward calculation shows that the multi-oscillator vector
potential operator satisfies, for any space-time points $x$, $y$,
$z$, the commutators 
\be
[{\uu A}_a(x),{\uu A}_b(y)]
&=&
\int d\Gamma(\bbox k)\uu I_{k}
\Big(
m_{a}(\bbox k)\bar m_b(\bbox k)e^{ik\cdot (y-x)}
-
\bar m_a(\bbox k)m_b(\bbox k)e^{ik\cdot (x-y)}
\Big)\nonumber
\\
&\pp =&+
\int d\Gamma(\bbox k)\uu I_{k}
\Big(
\bar m_a(\bbox k)m_b(\bbox k)e^{ik\cdot (y-x)}
-
m_{a}(\bbox k)\bar m_b(\bbox k)e^{ik\cdot (x-y)}
\Big)\label{[A,A]}\\
\big[[{\uu A}_a(x),{\uu A}_b(y)],{\uu A}_c(z)\big]
&=&0.
\ee
To obtain more insight as to the meaning of the commutator
(\ref{[A,A]}) consider its coherent-state average evaluated in a state
of the form (\ref{mocs}): 
\be
\langle \uu O_\alpha|
[{\uu A}_a(x),{\uu A}_b(y)]
|\uu O_\alpha\rangle
&=&
\int d\Gamma(\bbox k)Z(\bbox k)
\Big(
m_{a}(\bbox k)\bar m_b(\bbox k)e^{ik\cdot (y-x)}
-
\bar m_a(\bbox k)m_b(\bbox k)e^{ik\cdot (x-y)}
\Big)
\\
&\pp =&+
\int d\Gamma(\bbox k)Z(\bbox k)
\Big(
\bar m_a(\bbox k)m_b(\bbox k)e^{ik\cdot (y-x)}
-
m_{a}(\bbox k)\bar m_b(\bbox k)e^{ik\cdot (x-y)}
\Big).
\ee
The Minkowski-space metric tensor can be decomposed
\cite{PR} in terms of null tetrads as follows 
\be
g_{ab}
&=&
k_a\omega_b+\omega_a k_b-m_a\bar m_b-m_b\bar m_a.\label{g-id}
\ee
With the help of this identity we can write 
\be
\langle \uu O_\alpha|
[{\uu A}_a(x),{\uu A}_b(y)]
|\uu O_\alpha\rangle
&=&
\int d\Gamma(\bbox k)Z(\bbox k)
\big(k_{a}\omega_b(\bbox k)
+
k_{b}\omega_a(\bbox k)
\big)
\big(
e^{ik\cdot (y-x)}
-
e^{ik\cdot (x-y)}
\big)\nonumber\\
&\pp =&+g_{ab}
\int d\Gamma(\bbox k)Z(\bbox k)
\big(
e^{ik\cdot (x-y)}
-
e^{ik\cdot (y-x)}
\big)\label{<A>}
\ee
It is known that terms such as the first integral vanish if the
potential couples to a conserved current. The same property
guarantees gauge independence of the formalism. Therefore we can
concentrate only on the explicitly gauge independent term
proportional to $g_{ab}$. 
Denote 
\be
D_Z(x)
&=&
i\int d\Gamma(\bbox k)Z(\bbox k)
\big(
e^{-ik\cdot x}
-
e^{ik\cdot x}
\big)
\ee
For $Z(\bbox k)={\rm const}=Z$ we get $D_Z(x)$ proportional to the
Jordan-Pauli function,
\be
D_{Z}(x)=Z\, D(x)
\ee
which vanishes for spacelike $x$. However, the choice of
constant $Z(\bbox k)$ is excluded by the requirement of
square-integrability of $O$. Therefore the requirement that the
vacuum state be square-integrable seems to introduce some kind of
nonlocality into the formalism. 

There are two possibilities one can contemplate. First of all, one
can perform the calculations with arbitrary $O$ and then perform a
renormalization (we have seen that such a step is necessary even in
the nonrelativistic case). After the renormalization we can go to the
``flat" pointwise limit $Z(\bbox k)\to 0$, $\parallel O\parallel=1$, 
corresponding to the uniform
distribution of all the frequencies. Second, performing the
calculations in a preferred frame we can consider equal-time
commutation relations 
\be
\langle \uu O_\alpha|
[{\uu A}_a(t,\bbox x),{\uu A}_b(t,\bbox y)]
|\uu O_\alpha\rangle
&=&
\int d\Gamma(\bbox k)Z(\bbox k)
\big(k_{a}\omega_b(\bbox k)
+
k_{b}\omega_a(\bbox k)
\big)
\big(
e^{i\vec k\cdot (\vec y-\vec x)}
-
e^{i\vec k\cdot (\vec x-\vec y)}
\big)\nonumber\\
&\pp =&+g_{ab}
\int d\Gamma(\bbox k)Z(\bbox k)
\big(
e^{i\vec k\cdot (\vec x-\vec y)}
-
e^{i\vec k\cdot (\vec y-\vec x)}
\big)
\ee
The last term will vanish if 
\be
Z(\bbox k)=Z(-\bbox k)
\ee
i.e. if the vacuum is 3-inversion invariant. 
This can hold, however only in one reference frame unless $O$ is
constant, which we exclude. 

One can conclude that non-canonically quantized electrodynamics
is not a local quantum field theory, at least in the strict standard
sense. This is not very surprizing if one recalls that the
thermodynamic limit of nonrelativistic theories discussed in \cite{I}
and \cite{II} was equivalent to their canonical versions but with
cut-offs. The presence of $O$ in the integrals introduces some kinds of
effective extended structures, a consequence of nontrivial structures
of non-canonical vacua. 
The issue requires further studies. In particular, it is
important to understand an influence of the thermodynamic limit
$N\to\infty$ on locality problems in the context of relativistic
perturbation theory.  

There is an intriguing analogy between the kind of non-locality we
have obtained and the one encountered in field theory in
non-commutative space-time \cite{geometry}.

\section{Radiation fields associated with classical currents}

The problem of radiation fields is interesting for several reasons.
First of all, the radiation fields satisfy homogeneous Maxwell
equations so that the theory we have developed 
can be directly applied. Second, this is one of the simplest ways of
addressing the question of infrared divergences within the
non-canonical framework. 

It is widely known \cite{IZBB,BD} that  in the
canonical theory the scattering matrix corresponding
to radiation fields produced by a classical transversal 
current is given, up to a phase, by a
coherent-state displacement operator 
$e^{-i\int d^4y J(y)\cdot A_{\rm in}(y)}$. 
One of the consequences of such
an approach is the Poissonian statistics of photons emitted by
classical currents. An unwanted by-product of the construction is the infrared
catastrophe. 

Starting with the same $S$ matrix but expressed in terms of
non-canonical ``in" fields we obtain the non-canonical displacement
operator. 
Photon statistics is Poissonian in the thermodynamic limit but the infrared
catastrophe is automatically eliminated.

\subsection{Classical radiation field}

Let us assume that we deal with a classical transversal current $J_a(x)$
whose Fourier transform is $\tilde J_a(k)=\int d^4x e^{ik\cdot x}J_a(x)$. 
Transversality means here that 
\be
\tilde J_a(|\bbox k|,\bbox k)=
\bar m_a(\bbox k)\tilde J_{10'}(|\bbox k|,\bbox k) 
+ 
m_a(\bbox k)\tilde J_{01'}(|\bbox k|,\bbox k).
\ee
Formally, a solution of Maxwell equations
\be
\Box{\uu A}_a(x)=J_a(x)\uu I
\ee
can be written as 
\be
{\uu A}_a(x)
&=&
{\uu A}_{a\rm in}(x)+\int d^4y D_{\rm ret}(x-y)J_a(y)\uu I\\
&=&
{\uu A}_{a\rm out}(x)+\int d^4y D_{\rm adv}(x-y)J_a(y)\uu I.
\ee
Here ${\uu A}_{a\rm in}$ and ${\uu A}_{a\rm out}$ are solutions of
homogeneous equations. $D_{\rm ret}$ and $D_{\rm adv}$ are the
retarded and advanced 
Green functions whose difference is the Jordan-Pauli function 
\be
D(x)=i\int d\Gamma(\bbox k)\big(e^{-ik\cdot x}-e^{ik\cdot x}\big).
\ee 
The 4-potential of the radiation field is 
\be
{\uu A}_{a\rm rad}(x)
&=&
{\uu A}_{a\rm out}(x)
-
{\uu A}_{a\rm in}(x)=\int d^4y D(x-y)J_a(y)\uu I
\ee
and leads to the field spinors
\be
\uu\varphi_{XY\rm rad}(x)
&=&
\int d\Gamma(\bbox k)
\Big(\pi{_{(X}}\bar\pi{^{Y'}} \tilde J_{Y)Y'}(k)e^{-ik\cdot x}
+\pi{_{(X}}\bar\pi{^{Y'}}\overline{\tilde J}_{Y)Y'}(k)e^{ik\cdot x} \Big)
\uu I\\
{\uu{\bar\varphi}}_{X'Y'\rm rad}(x)
&=&
\int d\Gamma(\bbox k)
\Big(\bar\pi{_{(X'|}}\pi{^{Y}} \overline{\tilde J}_{Y|Y')}(k)e^{ik\cdot x}
+\bar\pi{_{(X'|}}\pi{^{Y}}\tilde J_{Y|Y')}(k)e^{-ik\cdot x} \Big)
\uu I.
\ee
Comparing these formulas with expressions (\ref{varphi}) and
(\ref{varphi'}) valid for all solutions of free Maxwell equations one
finds 
\be
f(\bbox k,+)
&=&
-\bar m^a(\bbox k)\tilde J_{a}(|\bbox k|,\bbox k)
=\tilde J_{01'}(|\bbox k|,\bbox k)=j(\bbox k,+)\label{J01}\\
f(\bbox k,-)
&=&
-m^a(\bbox k)\tilde J_{a}(|\bbox k|,\bbox k)
=\tilde J_{10'}(|\bbox k|,\bbox k)=j(\bbox k,-)\label{J10}\\
\uu a(\bbox k,s)_{\rm out}
&=&
\uu a(\bbox k,s)_{\rm in}+
j(\bbox k,s)\uu I.\label{in-out}
\ee
\subsection{Non-canonical radiation field}

Formula (\ref{in-out}) is analogous to the one from the canonical
theory. It is clear that although $\uu a(\bbox k,s)_{\rm in}$ and
$\uu a(\bbox k,s)_{\rm out}$ cannot be simultaneously of the form given by
(\ref{53}) and (\ref{58}), they do satisfy the non-CCR algebra 
(\ref{65}). In spite of this the result (\ref{in-out}) is
not very satisfactory. Indeed, one expects that the scattering matrix
describing a non-canonical quantum field interacting with a classical
current is
\be
S=e^{i\phi}e^{-i\int d^4y J(y)\cdot\uu A_{\rm in}(y)}\label{S-m}
\ee
with some phase $\phi$. Then 
\be
\uu A_{a\rm out}(x)
&=&
S^{\dag}\uu A_{a\rm in}(x)S\label{S-matrix}\\
&=&
\uu A_{a\rm in}(x)
-i\int d^4y J^{b}(y) 
[\uu A_{a\rm in}(x),\uu A_{b\rm in}(y)]\label{Arad}.
\ee
Employing (\ref{[A,A]}), (\ref{J01}), (\ref{J10}) one can write 
\be
\uu A_{a\rm rad}(x)
&=&
i\int d^4y J^{b}(y) \int d\Gamma(\bbox k)
\uu I_{\bbox k}
\Big(
\big(
e^{ik\cdot (x-y)}\bar m_am_b 
-e^{-ik\cdot (x-y)}m_a\bar m_b
\big)
+
\big(
e^{ik\cdot (x-y)}m_a\bar m_b
-e^{-ik\cdot (x-y)}\bar m_am_b
\big)
\Big)\\
&=&
i\int d\Gamma(\bbox k)
\uu I_{\bbox k}\Big(m_a\big(e^{-ik\cdot x}j(\bbox k,+)
-e^{ik\cdot x}\overline{j(\bbox k,-)}\big)
+
\bar m_a\big(e^{-ik\cdot x}j(\bbox k,-)
-e^{ik\cdot x}
\overline{j(\bbox k,+)}\big)
\Big),
\ee
where $m_a=m_a(\bbox k)$, and 
\be
\uu a(\bbox k,s)_{\rm out}
&=&
\uu a(\bbox k,s)_{\rm in}+
j(\bbox k,s)\uu I_{\bbox k}\label{in-out'}\\
&=&
\uu {\cal D}(j)^{\dag}
\uu a(\bbox k,s)_{\rm in}
\uu {\cal D}(j).
\ee
Consequently, the $S$ matrix is in the non-canonical theory
proportional to the {\it non-canonical\/} displacement operator
\be
S=e^{i\phi}\uu {\cal D}(j).
\ee
This fact will be shown to eliminate the infrared catastrophe. 

\subsection{Propagators}

Evaluating the average of (\ref{Arad}) in a coherent state
$|\uu O_\alpha\rangle$ one finds 
\be
\langle\uu O_\alpha|
\uu A_{a\rm rad}(x)
|\uu O_\alpha\rangle
&=&
\int d^4y D_Z(x-y) J_a(y)+{\rm gauge \,term}.\label{+gauge}
\ee
The irrelevant gauge term is a remainder of the first part of
(\ref{<A>}).  As expected the radiation field does not depend on what
$\alpha$ one takes in the coherent state, but does depend on the
vacuum structure. The presence of the regularized function 
$D_Z(x-y)$ instead of $D(x-y)$
implies that the radiation signal propagates in a neighborhood of the
light cone. Any deviations from $c$ in velocity of signal propagation
can be regarded as indications of a non-constant vacuum wave function
$O$. However, even in the orthodox canonical quantum 
electrodynamics a detailed analysis of signal propagation leads to
small deviations from velocity of light, especially at small
distances \cite{Man}. It may be difficult to experimentally
distinguish between the two effects. A similar effect was predicted
for Maxwell fields in non-commutative space-time \cite{Jackiw,Cai}. 

Using (\ref{g-id}) one can rewrite (\ref{[A,A]}) as 
\be
[{\uu A}_a(x),{\uu A}_b(y)]
&=&
g_{ab}
\int d\Gamma(\bbox k)\uu I_{\bbox k}
\Big(e^{ik\cdot (x-y)}-e^{ik\cdot (y-x)}\Big)
+\dots\label{hat D}
\ee
where the dots stand for all the terms which are gauge dependent and
do not contribute to physically meaningful quantities. 
We can therefore identify 
\be
\uu D(x)=i\int d\Gamma(\bbox k)\uu I_{\bbox k}
\Big(e^{-ik\cdot x}-e^{ik\cdot x}\Big)
\ee
as the operator responsible for the appearence of the smeared out
Jordan-Pauli function $D_Z$ in the
coherent-state average (\ref{+gauge}). $\uu D(x)$ is a
translation-invariant scalar-field operator solution of the
d'Alambert equation, i.e. 
\be
\Box \uu D(x)&=&0,\label{d'A}\\
\uu V_{\Lambda,y}^{\dag}\uu D(x) \uu V_{\Lambda,y}&=&\uu
D(\Lambda^{-1}x). 
\ee
The operator analogues 
of retarded and advanced Green functions are  
\be
\uu D_{\rm ret}(x)
&=&
\Theta(x_0)\uu D(x),\\
\uu D_{\rm adv}(x)
&=&
-\Theta(-x_0)\uu D(x),\\
\uu D(x)&=&\uu D_{\rm ret}(x)-\uu D_{\rm adv}(x).
\ee
Eq. (\ref{d'A}) implies that the operators 
\be
\int d^4y \uu D_{\rm ret}(x-y)J_a(y)
\ee
and 
\be
\int d^4y \uu D_{\rm adv}(x-y)J_a(y)
\ee
differ by at most a solution of the homogeneous equation
\be
\Box \uu A_a(x)=0. \label{hom}
\ee
(\ref{d'A}) implies also that one can define 
\be
\uu \delta(x)=\Box \uu D_{\rm adv}(x)=\Box \uu D_{\rm ret}(x).
\ee
It follows that having a solution ${\uu A}_{a\rm in}(x)$ of
(\ref{hom}) one can define another solution ${\uu A}_{a\rm out}(x)$ of
(\ref{hom}) by means of
\be
{\uu A}_a(x)
&=&
{\uu A}_{a\rm in}(x)+\int d^4y \uu D_{\rm ret}(x-y)J_a(y)\\
&=&
{\uu A}_{a\rm out}(x)+\int d^4y \uu D_{\rm adv}(x-y)J_a(y),
\ee
simulatneously guaranteeing that the correct $S$-matrix conditions 
(\ref{S-m}), (\ref{S-matrix}) are fulfilled up to, perhaps, a gauge
transformation. ${\uu A}_a(x)$ is a solution of 
\be
\Box \uu A_a(x)=\uu J_a(x) \label{inhom}
\ee
where 
\be
\uu J_a(x)&=&\int d^4y \uu \delta(x-y)J_a(y).
\ee
\subsection{The problem of infrared catastrophe}

To close the discussion let us consider the issue of infrared
catastrophe. We have to compute an avarage number of photons in the
state $\uu D(j)|\uu O\rangle$. The number-of-photons operator is 
\be
\uu n=\oplus_1 (1\otimes \sum_sa^{\dag}_sa_s).
\ee
The $1$ in the above formula is the identity in the $\bbox k$ space
and $a_s$ satisfy CCR. 
The average reads
\be
\langle n\rangle
&=&
\langle \uu O_j|\uu n|\uu O_j\rangle
=
\sum_s\int d\Gamma(\bbox k)Z(\bbox k)|j(\bbox k,s)|^2.
\ee
The four-momentum of the radiation field is 
\be
\langle P_a\rangle
&=&
\langle \uu O_j|\uu P_a|\uu O_j\rangle
=
\sum_s\int d\Gamma(\bbox k)k_a Z(\bbox k)|j(\bbox k,s)|^2.
\ee
$O(\bbox k)$ belongs to a carrier space of an
appropriate unitary representation of the Poincar\'e group. As such this is a
differentiable function vanishing
at the origin $k=0$ of the light cone. This is a consequence of the
fact that the cases $k=0$ and $k\neq 0$, $k^2=0$, correspond to
representations of the Poincar\'e group induced from $SL(2,C)$ and $E(2)$,
respectively (for another justification see \cite{Woodhouse}). 

Hence, the regularization of the infrared divergence is implied by
relativistic properties of the field. It is quite remarkable that all
the divergences are regularized automatically by the
same property of the formalism: The nontrivial structure of the
vacuum state. In the case of ultraviolet and vacuum divergences the
regularization is a consequence of square integrability of $O$. 

\section{On the meaning of $Z_3$}

Consider the class of vacua whose probability densities 
$Z(\bbox k)$ are constant over some region in
the $\bbox k$-space, say $Z(\bbox k)=Z$ for 
$k_{\rm min}<|\bbox k|<k_{\rm max}$, and decay to zero outside of
this plateau region. For $Z(\bbox k)$ in the class we find 
\be
\langle n\rangle
&\approx &
Z\sum_s\int_{\rm cut-off} d\Gamma(\bbox k)|j(\bbox k,s)|^2\\
\langle P_a\rangle
&\approx &
Z\sum_s\int_{\rm cut-off} d\Gamma(\bbox k)k_a |j(\bbox k,s)|^2.
\ee
The index ``cut-off" means restricting the integral to the plateau region
of $Z$. 
Although the
pointwise limit $Z(\bbox k)\to 0$ does not belong to the class, one
can nevertheless consider the  limits (in general of
the $0\cdot\infty$ type) 
\be
\langle n\rangle_{Z=0}
&=&
\lim_{Z\to 0,\rm pointwise}
\sum_s\int d\Gamma(\bbox k)Z|j(\bbox k,s)|^2,\label{Z1a}\\
\langle P_a\rangle_{Z_3=0}
&=&
\lim_{Z\to 0,\rm pointwise}
\sum_s\int d\Gamma(\bbox k)k_a Z|j(\bbox k,s)|^2,\label{Z1b}
\ee
which are essentially the formulas for renormalized quantities from
the standard formalism if one identifies $Z$ with the renormalization
constant $Z_3$. Actually, all the above formulas involving $Z(\bbox
k)$ contain $Z$ in exactly those places where one expects $Z_3$ to
appear in renormalized quantities \cite{IZBB}. 
The (smooth) cut-offs appear 
automatically through the asymptotic properties of $Z(\bbox k)$ at zero
and infinity. One can thus speak of (finite!) self-renormalization of the
non-canonical theory. 
Let us note that both perturbative and nonperturbative
calculations presented in \cite{I} and \cite{II} were pointing 
into a reinterpretation of $Z(\bbox k)^\frac{1}{2}e_0$ 
in terms of an effective physical
charge. 

The formulas (\ref{Z1a}), (\ref{Z1b}) have to be
supplemented by 
\be
\lim_{Z\to 0,\rm pointwise}
\int  d\Gamma(\bbox k) Z= 
\int d\Gamma(\bbox k) Z(\bbox k)=1.
\ee
The fact that the pointwise limit gives predictions of the
renormalized theory suggests that the physical vacuum is indeed
``flat" and $Z=Z_3\approx 0$. The interpretation suggests also that the
value and even a functional form of $Z$ may depend on the state of
the entire Universe and, hence, change during its evolution. Now,
since 
\be
\alpha=\frac{e^2}{\hbar c}=\frac{e_0^2Z_3}{\hbar c}
\ee
one may may treat the observed astrophysical changes of $\alpha$
\cite{alpha} as an indication of evolving $Z(\bbox k)$. 

\section{Conclusions and further perspectives}

We have formulated a manifestly covariant version of non-canonically
quantized electromagnetic fields. 
A thermodynamic limit of non-canonical
theories looks like a {\it finite\/} canonical theory, explaining the
success of the latter. In the limit of an ``infintely flat" vacuum
(meaning the cut-off at infinity and $Z\to 0$) the theory becomes
local. The elements such as ultraviolet and
infrared formfactors do not have to be introduced in ad hoc manners
but follow from the more fundamental non-canonical level. 

An issue which has not been addressed so far is how to quantize
fermions. Some results on the Dirac equation 
are known already and will be presented in a
separate paper. Another problem is to embed the concrete non-canonical
quantization procedure we have proposed into a more abstract scheme
of quantizations in a $C^*$-algebraic setting. The fact that the
right-hand-side of commutation relations is not an identity but
rather an operator belonging to the center of the algebra suggests
directions for generalizations. It seems there is a link with the
work of Streater on non-abelian cocycles \cite{Streater}. 
An appropriate version
of a coherent-state quantization based on the formalism of Naudts and
Kuna \cite{NK} is in preparation. 

\acknowledgments
The work was done mainly during my stays in Antwerp and Clausthal
with NATO and Alexander-von-Humboldt fellowships. I
am indebted to Prof. H.-D. Doebner, W. L\"ucke, J. Naudts and M. Kuna for many
interesting discussions.


\begin{references}
\bibitem{PT}In this paper when we speak of the Poincar\'e group we
mean the semidirect product of 4-translations and $SL(2,C)$, i.e. the
universal covering space of the Poincar\'e group.
\end{references}
\end{document}